\documentclass[twocolumn,preprintnumbers,amsmath,prb,amssymb]{revtex4}
\usepackage{graphicx}
\usepackage{dcolumn}
\usepackage{bm}
\usepackage[usenames,dvipsnames]{color}

\def \be {\begin{equation}}
\def \ee {\end{equation}}
\def \ben {\begin{eqnarray}}
\def \een {\end{eqnarray}}
\def\re#1{{\color{black} {#1}}}

\begin{document}


\title{Effects of Donor-Acceptor Quantum Coherence and Non-Markovian Bath on the Distance Dependence of Resonance Energy Transfer}

\author{Seogjoo J. Jang}
\affiliation{Department of Chemistry and Biochemistry, Queens College, City University of New York, 65-30 Kissena Boulevard, Queens, New York 11367\footnote{mailing address}  \& PhD programs in Chemistry and Physics, and Initiative for the Theoretical Sciences, Graduate Center, City University of New York, 365 Fifth Avenue, New York, NY 10016}
\email{SJANG@QC.CUNY.EDU}

\date{Published in the {\it Journal of Physical Chemistry C} {\bf 123}, 5767-5775 (2019) }
\begin{abstract}
Accurate information on the distance dependence of resonance energy transfer (RET) is crucial for its utilization as a spectroscopic ruler \re{of} nanometer scale distances.  In this regard, understanding the effects of donor-acceptor quantum coherence and non-Markovian bath, which become significant at short distances, has significant implications. The present work investigates this issue theoretically by comparing results from a theory of coherent RET (CRET) with a nonequilibrium version of F\"{o}rster's RET (FRET) theory, both accounting for non-Markovian bath effects.  Even for a model where the donor-acceptor electronic coupling is of transition dipole interaction form, it is shown that the  RET rate in general deviates from the inverse sixth power distance dependence as opposed to the prediction of the original FRET.   It is shown that the donor-acceptor quantum coherence makes the \re{distance} dependence steeper than the sixth power although detailed manner of enhancement is sensitive to specific values of parameters.  On the other hand, the non-Markovian bath effects make the \re{distance} dependence more moderate than the sixth power for both CRET and  nonueqilibrium FRET because finite time scale of the bath causes the rate to be  smaller than the prediction of original FRET.  While these effects are \re{demonstrated clearly} in the population dynamics at sub-picosecond time scales, 
their contributions to the conventional RET efficiency are relatively minor.  This indicates that the actual detection of such effects through conventional RET efficiency measurement requires either high precision or utilization of a donor with fast spontaneous decay rate of excitation.    

\end{abstract}

\maketitle

\section{Introduction}
F\"{o}rster's resonance energy transfer (FRET),\cite{forster-ap,forster-dfs,forster-book-1} the theory that laid a fundamental quantum mechanical basis for electronic excitation transfer processes in general,\cite{silbey-arpc27,ret,nitzan,scholes-arpc54,olaya-castro-irpc30,jang-wires3,jang-rmp90} is also well known for its application for nanometer scale distance measurement.\cite{stryer-pnas58,roy-nm5,selvin-nsb7,sahoo-jppc12,heyduk-cob13,schuler-cosb18}  The utility of FRET as a spectroscopic ruler comes from the fact that its efficiency,
\be
E=\frac{1}{1+(R/R_0)^6}  , \label{eq:fret_eff}
\ee
is sensitive to the distance $R$ between the donor ($D$) and the acceptor ($A$) of excitation near the F\"{o}rster radius $R_0$, typically in the range of $2-10 \ {\rm nm}$ (see Appendix A for the definition of $R_0$ and the derivation of eq. \ref{eq:fret_eff}).   Various versions of FRET efficiency measurements are now well established, and the areas of applications employing FRET continue growing.\cite{jares-erijman-nb21,zhang-acie50,berezin-cr110,piston-tbs32,zhang-acsnano6,das-jcp135,preus-cbc13,roda-abc393,hillisch-cosb11,heyduk-cob13,ha-arpc63,lee-me581,dimura-cosb40,dagher-nn13,melle-jpcb122}    However, despite decades of advances, the promise of FRET  as a genuine spectroscopic ruler with reliable precision has not been realized yet.  FRET efficiency
\re{as a quantitative probe} for structural change, in particular at single molecule level,\cite{ha-pnas93,roy-nm5,heyduk-cob13,weiss-science283,schuler-cosb18,guo-pccp16,basak-pccp16,chung-pccp16,stockmar-jpcb120} \re{is} now well established, but \re{it remains challenging to gain}
accurate \re{enough} distance information despite various efforts.\cite{ha-pnas93,ha-arpc63,lipman-science301,schuler-pnas102,schuler-cosb18,nir-jpcb110,watkins-jpca110,dimura-cosb40,gansen-nc9}  Thus, \re{improving the utility} of FRET, or more generally RET,\cite{bates-prl94}  for more precise distance determination remains an important issue,\cite{kalinin-nm9,hellenkamp-nm15} which in turn requires better understanding and control of \re{the} distance dependence \re{of RET}.  

The theory of FRET is based on the assumption of transition dipole interaction and the Fermi's golden rule (FGR). 
Equation \ref{eq:fret_eff} is the direct outcome of these assumptions and \re{an additional set of assumptions on the kinetics} as detailed in the Appendix A. 
\re{In reality}, many mechanisms\cite{scholes-arpc54,olaya-castro-irpc30,jang-wires3,jang-rmp90,khan-cpl461,munoz-losa-bj96,scholes-jpcb111,swathi-jcp130,ortiz-jpcb109,fuckel-jcp128,langhals-jacs132,metivier-prl98,nalbach-prl108,dung-pra66,wubs-njp18,hsu-jpcl8,du-cs9}  
can complicate \re{such approximations and} assumptions.   Thus, the deviation of the distance dependence of RET efficiency from eq. \ref{eq:fret_eff} in real systems is not unexpected.   \re{The simple case where this is caused by a reverse FRET reaction is described in Appendix B.} 

\re{While there were some theoretical works addressing 
the distance dependence of RET}, 
most of them  still assumed eq. \ref{eq:fret_eff} as the starting point and explained experimental observations in terms of fluctuations\cite{makarov-jcp132,dolghih-jppa190,ortiz-jpcb109,badali-jcp134,yang-jcp121,platkov-jcp141,hummer-jpcb121,andryushchenko-jpcb122,gopich-jpcb107,gopich-pnas109,chung-pccp16} or distributions\cite{tanaka-jcp109,wang-jpcb109,ahn-cpl446,sahoo-jacs128,allen-jpc131,hilczer-jcp130,melle-jpcb122} of $R$  or orientations.   Such approach, while appropriate for many cases, would lead to incorrect assessment of the fluctuation/distribution of distances if the underlying mechanism deviates from the original FRET.  Therefore, establishing the correct distance dependence of the RET efficiency at \re{the level of individual RET process, prior to statistical averaging, is important for both qualitatively correct and quantitatively reliable modeling of experimental data.}

Within the approximation of the FGR, understanding how the distance dependence deviates from the original FRET is straightforward, which requires consideration of electronic couplings\cite{jang-rmp90,dexter-jcp,kuhn-jcp53,baer-jcp128,khan-cpl461,hsu-acr42,vura-weis-jpcc114} only.  The most well-known cases are when the electronic coupling between $D$ and $A$ involves higher order multipolar\cite{dexter-jcp,kuhn-jcp53,baer-jcp128,khan-cpl461} or exchange\cite{dexter-jcp} interaction. The distance dependence in each of these cases becomes a polynomial of $1/R$ or exponential function of $R$.  Less known mechanisms that have been clarified more recently are multichromophoric\cite{jang-prl92} and inelastic\cite{jang-jcp127,langhals-jacs132,nalbach-prl108} FRET. \re{Although somewhat in different context,  recent studies showed that nonadiabatic effects\cite{chin-np9,tiwari-pnas110,tiwari-jcp147,peters-jcp147,hestand-cr118} can also make significant contribution.} For these, the distance dependences are less apparent because of interactions between  electronic couplings and multiple electronic/vibrational states.   Typically, the effective electronic coupling constants for these become sensitive to temperature as well.\cite{langhals-jacs132,nalbach-prl108}

If the electronic coupling between $D$ and $A$ is large compared to other energetic parameters, the \re{assumption behind }  FGR 
breaks down and it becomes necessary to consider the $D-A$ quantum coherence and non-Markovian/nonequilibrium bath effects.  How these 
re{factors} alter the distance dependence is an issue that has received relatively little attention, and is the main focus of the present work. \re{To this end,} 
explicit calculation of population dynamics is necessary 
\re{, unlike the cases that use FGR-based rates.}  Section II describes the model and method of such calculation being employed here, and Sec. III provides results of population dynamics and shows dependences of resulting RET rates on the distance.  Section IV discusses the implications of these results for the measurement of RET efficiency, and Sec. V offers a conclusion along with issues to be investigated further in the future.

\section{Theoretical models and computational methods}

Because the term coherence can be used in many different contexts, it is useful to first make its definition clear.   The $D-A$ quantum coherence considered here is defined as broadly as possible, and represents all the cases where the excitation transfer dynamics cannot be described by the first order time dependent perturbation theory, regardless of whether actual coherent population dynamics in time are observed or not.    It is important to note that the presence or observation of $D-A$ quantum coherence also depends on specifics of the initial condition and the relative excitation energies of $D$ and $A$.

In general, the effects of $D-A$ quantum coherence are significant at short distances, where other mechanisms causing deviation from FRET can also be substantial.  Thus, in real systems, simultaneous consideration of all such important factors would be necessary.  Nonetheless, in order to focus on the effects of quantum coherence only, it is assumed here that the $D-A$ electronic coupling is of purely transition dipole interaction form at all distances.  Thus, the donor-acceptor coupling $J$ is assumed to be 
\be
 J=J_0\left(\frac{R_0}{R}\right )^3 ,\label{eq:j_def}
 \ee
where $J_0$ is the value of electronic coupling at $R=R_0$, the F\"{o}rster radius, and is assumed to be independent of time.   Various relationships between $J_0$ and other parameters are detailed in the Appendix A.
    
A \re{minimal} model is used here.  The excitation energy of $D$ is denoted as $E_D$ and that of $A$ is denoted as $E_A$.  Thus, the electronic part of the Hamiltonian in the single exciton space is given by 
\be
H_e=E_D|D\rangle\langle D|+E_A|A\rangle\langle A|+J(|D\rangle\langle A|+|A\rangle\langle D|)\ ,
\ee   
 where $|D\rangle$ represents the state that only $D$ is excited and $|A\rangle$ the state that only $A$ is excited.  The ground electronic state and the spontaneous decay to this is not explicitly considered here.  All the degrees of freedom coupled to the two excited electronic states are designated collectively as the bath and are assumed to be modeled by harmonic oscillators linearly coupled to $|D\rangle$ and $|A\rangle$.  Thus, the total Hamiltonian is given by $H=H_e+H_{eb}+H_b$, where \re{$H_{eb}=\sum_n \hbar\omega_n (b_n+b_n^\dagger)(g_{nD}|D\rangle\langle D|+g_{nA}|A\rangle\langle A|)$} and $H_b=\sum_n\hbar\omega_n (b_n^\dagger b_n+1/2)$.
The spectral density for $D$ or $A$ is defined as 
\begin{equation}
\mathcal{J}_{D/A}(\omega)=\pi\hbar \sum_{n}\delta (\omega-\omega_{n})\omega_{n}^{2}g_{n_{D/A}}^{2} .
\end{equation}
For simplicity, it is assumed that there are no common bath modes\cite{beljonne-jpcb113,hennebicq-jcp130} coupled to both $D$ and $A$ although they may have significant role at short distances. For all the model calculations conducted here, the following super-Ohmic spectral density is assumed.  
\be
{\mathcal J}_{_{D/A}}(\omega)=\pi\hbar\frac{\eta_{_{D/A}}}{3!}\frac{\omega^3}{\omega_c^2} e^{-\omega/\omega_c} \  . \label{eq:super-ohmic}
\ee
\re{This is one of the most frequently used model spectral densities\cite{jang-rmp90}  to represent the combined response of molecular vibrations and environmental response upon electronic excitation. For this spectral density, the reorganization energy is $\hbar \eta \omega_c/3$ and the maximum value occurs at $\hbar\omega_c/3$.  Thus, it is reasonable to choose $\omega_c$ to represent the dominant molecular vibrational mode coupled to the excitation.}

\begin{figure}
\includegraphics[width=3.5in]{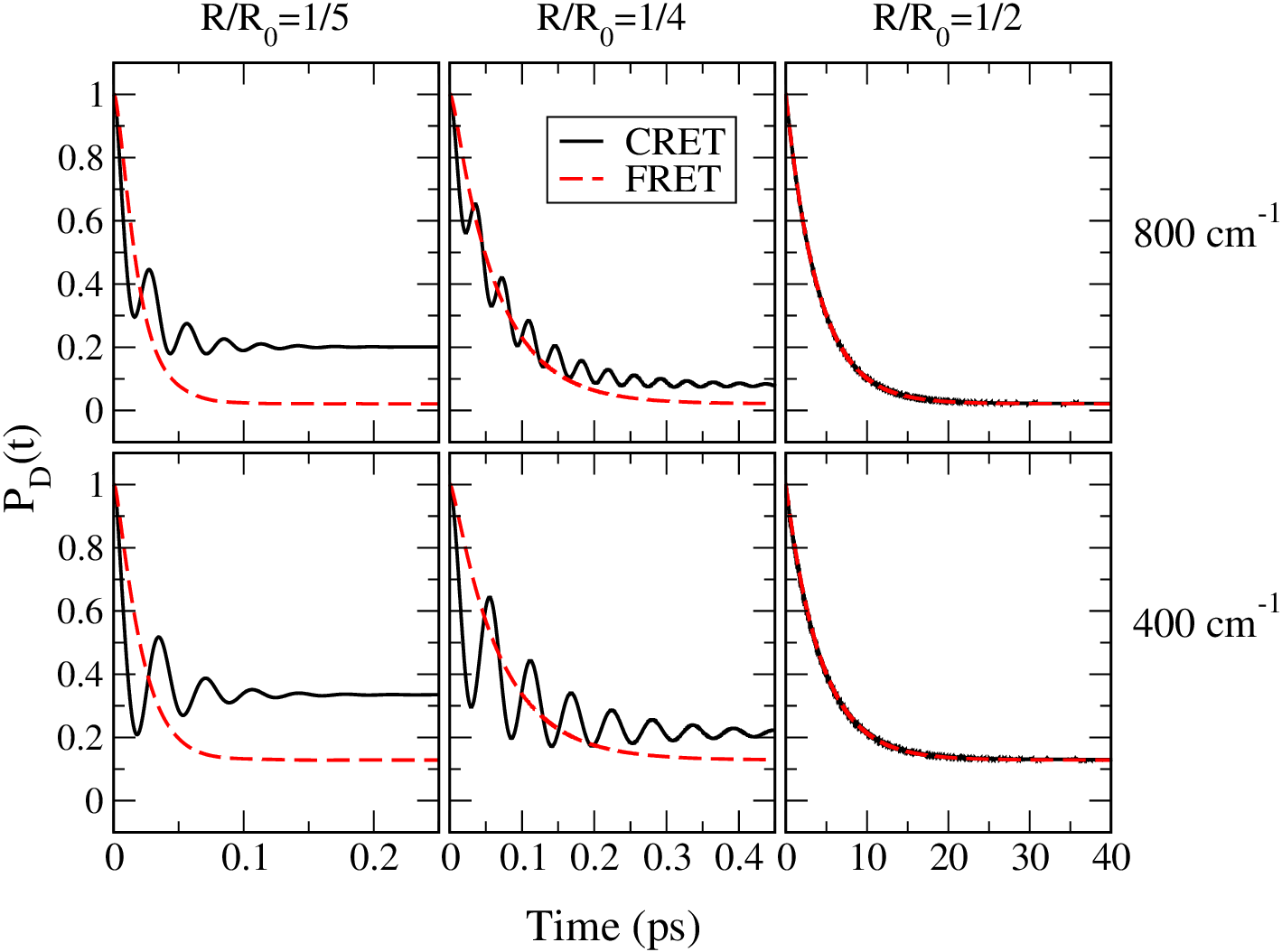}
\caption{Excited $D$ populations \re{for $\eta_D=\eta_A=2$, $\hbar \omega_c=1,000\ {\rm cm^{-1}}$ (Case I)} for two different values of $E_D-E_A=800$ and $400\ {\rm cm^{-1}}$  (shown on the right on each row), at three values of $R/R_0$ (shown on the top).  For all the cases, $J_0=5\ {\rm cm^{-1}}$ and $T=300\ {\rm K}$.  Black solid lines represent results from CRET and red dashed lines from FRET. }
\end{figure}

\begin{figure}
\includegraphics[width=3.5in]{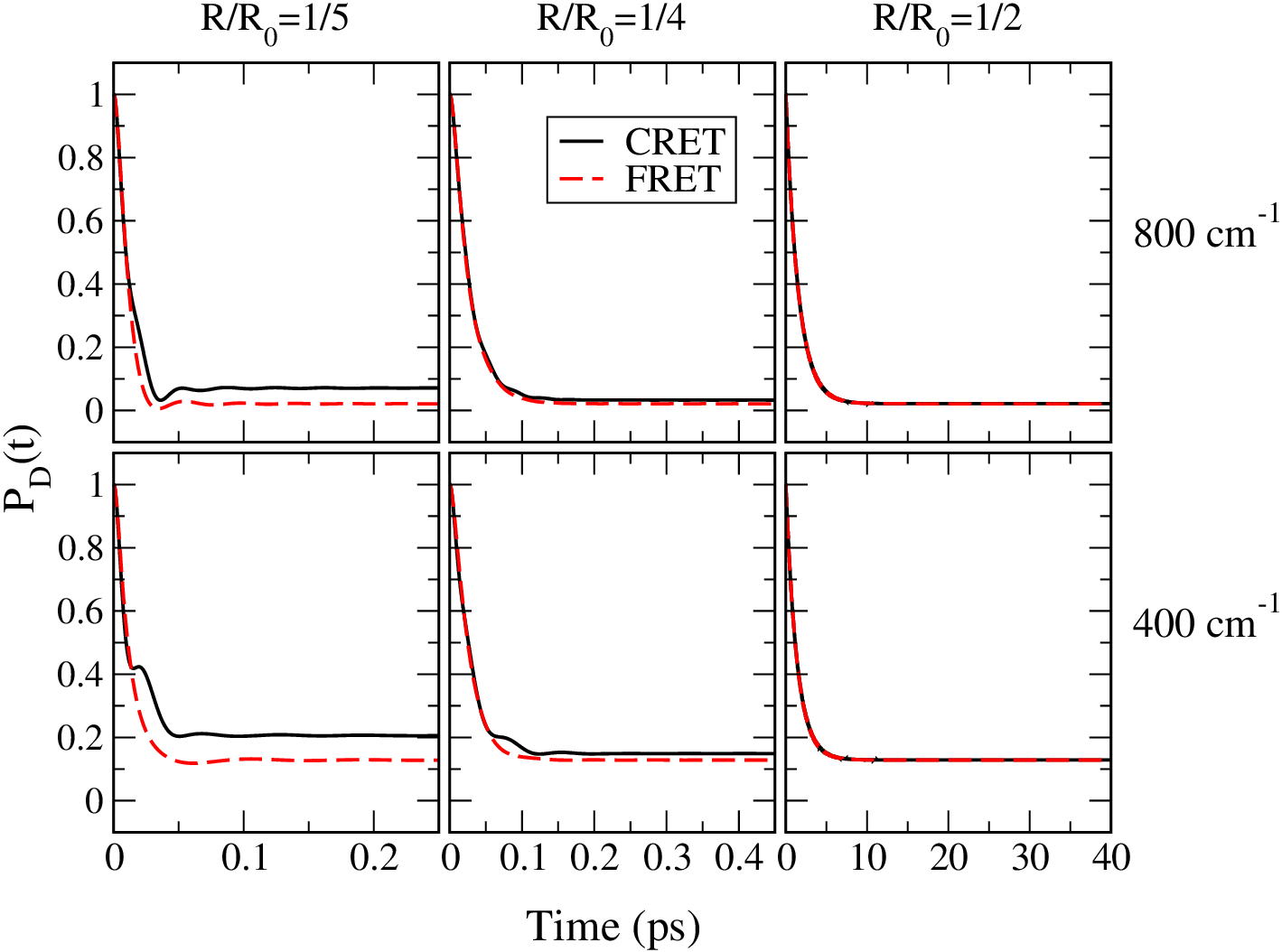}
\caption{Excited $D$ populations for  $\eta_D=\eta_A=5$ and $\hbar \omega_c=400\ {\rm cm^{-1}}$ \re{(Case II)}. Other parameters and conditions are the same as Fig. 1. }
\end{figure}

The method of calculation being employed here is the theory of coherent RET (CRET) based on the second order approximation of polaron-transformed quantum master equation approach.\cite{jang-jcp131}   \re{While this is not exact, the polaron transformation employed here makes it possible to account for part of the exciton-bath coupling to an infinite order such that the resulting equation approaches both coherent weak system-bath coupling and  incoherent hoping regimes properly.}  For comparison, master equation calculation using time dependent FRET rate, which includes nonequilibrium effect,\cite{jang-jcp131,jang-cp275} is conducted as well.

\section{Results}

For $J_0=5\ {\rm cm^{-1}}$ and at $T=300 \ {\rm K}$, both CRET and nonequilibrium FRET populations were calculated as detailed in Ref. \onlinecite{jang-jcp131}.     
\re{A wide range of parameters for the spectral density of eq. \ref{eq:super-ohmic} have been tested, but only two representative cases with different time scales of the bath while having the same reorganization energy are presented here.   These two cases correspond to $\eta_D=\eta_A=2$ and $\omega_c/(2\pi c)=1,000\ {\rm cm^{-1}}$ (Case I) and to $\eta_{D}=\eta_A=5$ and $\omega_c/(2\pi c)=400\ {\rm cm^{-1}}$ (Case II).}

\re{Figure 1 shows results for Case I} at two values of $E_D-E_A=800$ and $400\ {\rm cm^{-1}}$.   For $R/R_0=1/5$, where $J=625\ {\rm cm^{-1}}$, the populations based on CRET exhibit strongly coherent behavior and approach steady state values that are significantly different from those of FRET.  This deviation indicates that the excitation in the steady state limit becomes substantially delocalized between $D$ and $A$, which \re{should} also alter the oscillator strengths and lifetimes of excited states.  For $R/R_0=1/4$, where $J=320\ {\rm cm^{-1}}$, the extent of population oscillation for CRET is similar, but its average is closer to that of FRET.  Finally, for $R/R_0=1/2$, where $J=40\ {\rm cm^{-1}}$, the coherence is virtually absent and the CRET population almost agrees with that of FRET.

\re{Figure 2 shows results for Case II} while other parameters and conditions remain the same as in Fig. 1.  \re{While both Case I and Case II have the same reorganization energy, the time scale of the bath for the latter is much longer.  On the other hand, due to larger value of $\eta$ for Case II }, the effects of quantum coherence are less apparent than those shown in Fig. 1.   The differences between CRET and FRET populations are relatively minor even for $R/R_0=1/5$.   As yet, the population dynamics clearly deviate from \re{the} exponential behavior at short distances for both CRET and FRET as was observed experimentally.\cite{das-jcp135}  This is caused by the non-Markovian and nonequilibrium effects of the bath, which become more significant in this case due to slower time scale \re{(smaller value of $\omega_c$)} of the bath.  \re{In other words, the slower the bath, it takes more time for the resonance condition expected from spectral overlap to be fully realized.}

Even though the population dynamics are non-exponential, it is useful to define an effective RET rate as follows: 
\be
k_{eff}=\frac{P_A(\infty)}{\tau_{_{RET}}} \label{eq:k_eff} ,
\ee
where $\tau_{_{RET}}$ is the shortest time that satisfies the condition, $P_D(\tau_{_{RET}})-P_A(\tau_{_{RET}})P_D(\infty)/P_A(\infty)=1/e$.  This
determines $\tau_{_{RET}}$ as the first time when the population of excited $D$ becomes $1/e$ factor of its initial condition \re{as has been shown in previous works\cite{jang-prl113,jang-jpcl6,jang-jpcl9} and also detailed in Appendix B}. The rate defined by eq. \ref{eq:k_eff} is the corresponding forward rate that satisfies the condition of detailed balance.   Figure 3 compares actual time dependent populations with exponential ones corresponding to $k_{eff}$, which shows that the latter captures the major time scale of transfer and also reproduces correct steady state limits.

\begin{figure}
\includegraphics[width=3.5in]{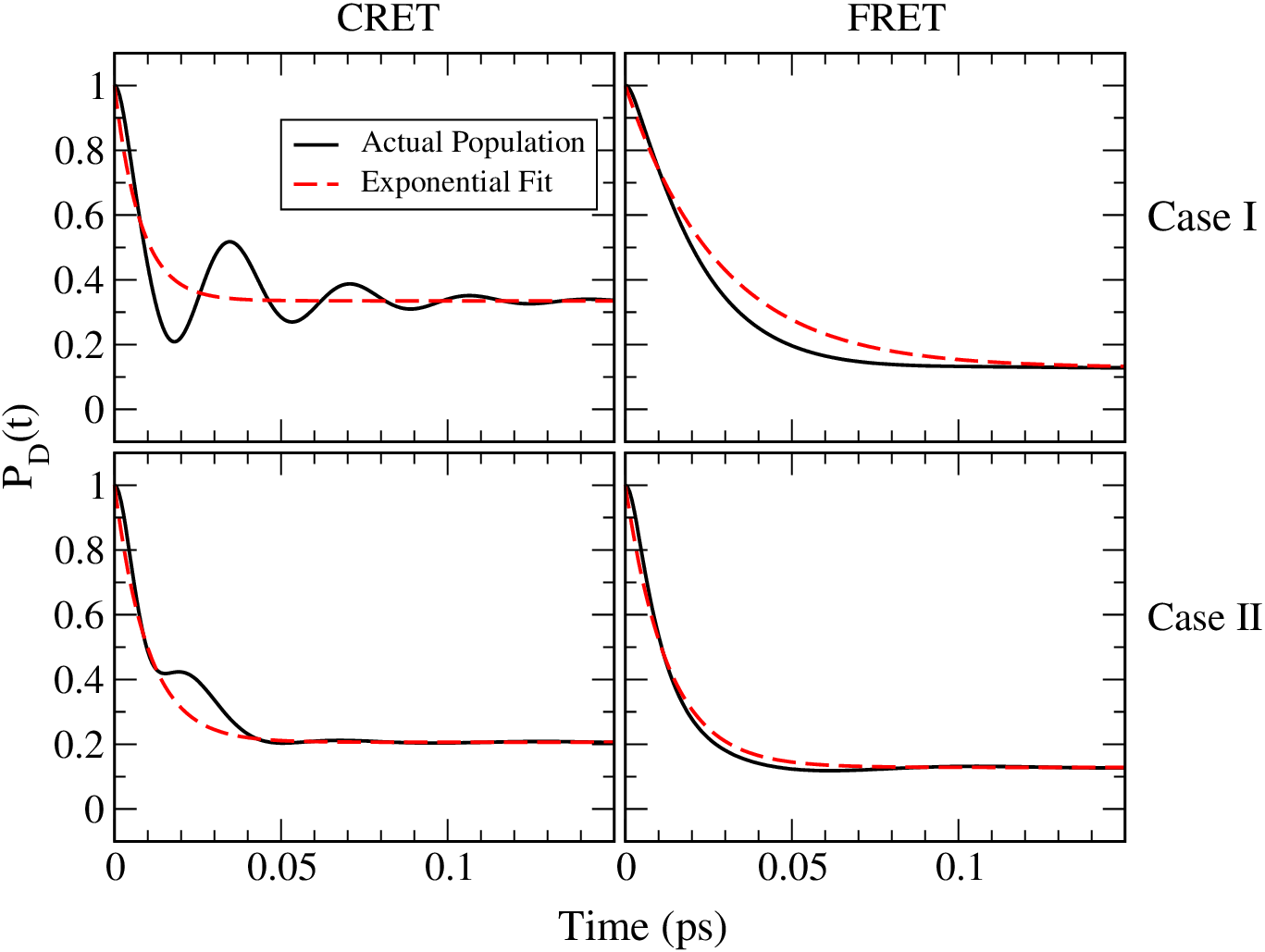}
\caption{Comparison of actual population dynamics based on either CRET or FRET with exponential dynamics corresponding to the effective rate, eq. \ref{eq:k_eff}, for $R/R_0=1/5$ and $E_D-E_A=400\ {\rm cm^{-1}}$. }
\end{figure}

Effective rates were determined at $1,000$ values of $R/R_0$ in the range of $1/5$ and $1/2$, by conducting explicit CRET and FRET calculations and using eq. \ref{eq:k_eff}.  Figure 4 provides the results of calculation.  The upper panels show the rates in logarithmic scale and the lower panels provide the same rates multiplied by $(R/R_0)^6$, which demonstrates the distance dependence of rates more clearly. The results for Case I show that coherent quantum dynamics results in significant enhancement of the transfer when compared to FRET.   The step-wise increase of $k_{eff}$ for short distances here is due to the specific definition of $\tau_{RET}$.   \re{This is because the enhancement in the oscillation rate of the donor population does not necessarily lead to significant change of $\tau_{RET}$ as long as the local minimum of the oscillation is above the threshold value, the $1/e$ factor of the initial value.  However,}  once averaged over fluctuations and also finite pulse width, such pattern is likely to disappear.    

The dependence of the FRET rates on inverse distance for Case I in Fig. 4, according to lower panels, is more moderate than $1/R^6$ 
\re{for} short distances.  For 
Case II, this trend is more pronounced.  The physical reason for this is the slowing down of excitation transfer due to non-Markovian effects of the bath, the relative proportion of which increases  as the distance becomes smaller.    

In summary, the results shown in Fig. 4 clarify two opposing effects.  First, the quantum coherence tends to make the dependence on inverse distance steeper than the original prediction of FRET.  This does not yet include the effect due to change in the oscillator strength and thus the spontaneous decay rate that can be significant when the excitation is delocalized between the donor and the acceptor.   Second, the nonequilibrium and non-Markovian effects lessen the dependence on inverse distance.  \re{This is because the delay due to finite time scale of the bath becomes relatively more significant as the electronic coupling becomes larger with the decrease of the $D-A$ distance.   It is also worthwhile to mention here that the extent of nonequilibrium effect depends on both the non-Markovian effect and the initial condition of the excited donor.   In this sense, the non-Markovian effect can be viewed as playing more fundamental role for delaying the RET dynamics. }

\begin{figure}
\includegraphics[width=3.5in]{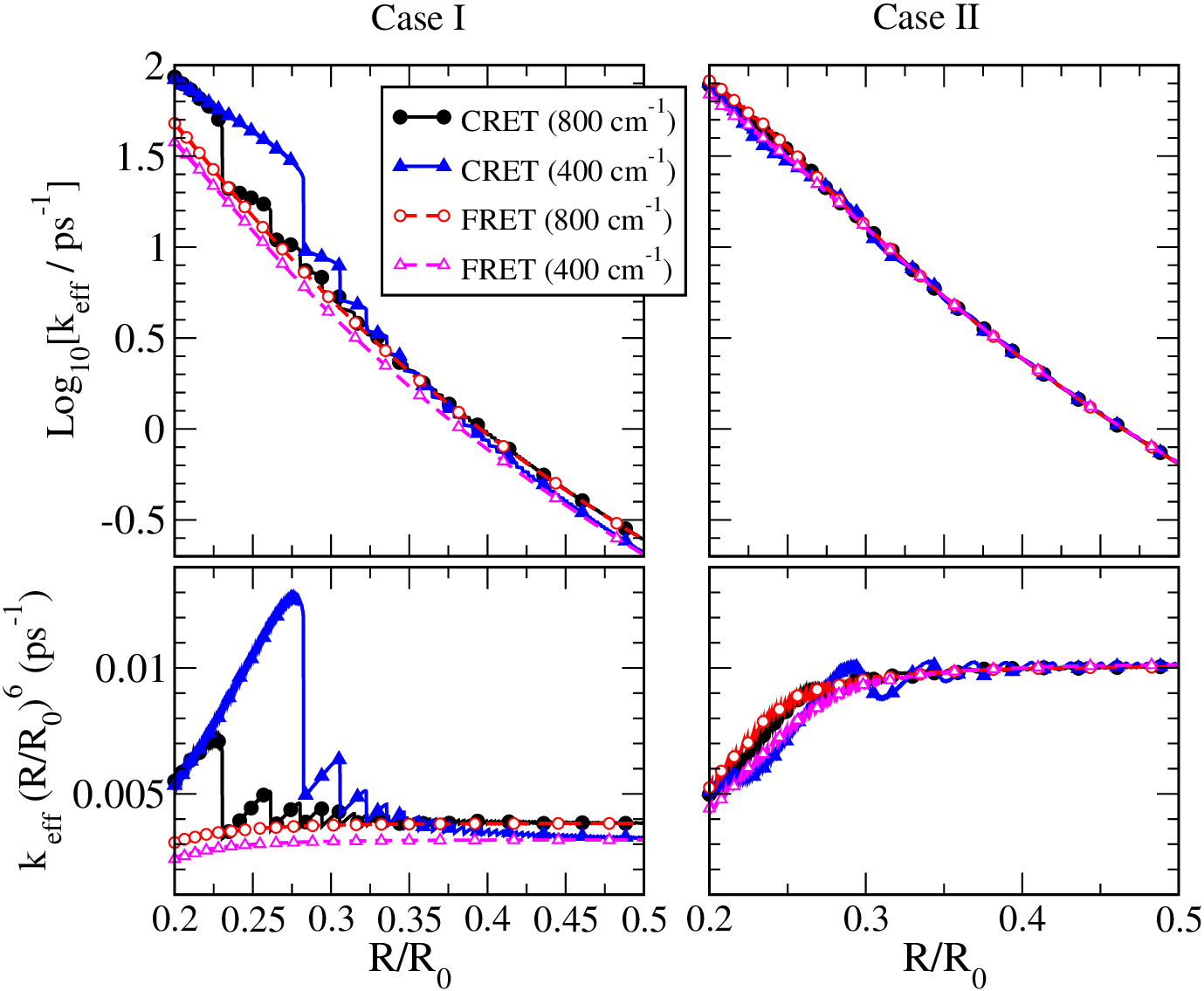}
\caption{Upper Panels: Effective rates ($k_{eff}$) in logarithmic scale versus with $R/R_0$.  Lower Panels:  Effective rates multiplied by $(R/R_0)^6$ versus $R/R_0$. }
\end{figure}

\section{Discussion}

Considering that all the FRET and CRET rates agree very well with each other already at $R/R_0=1/2$,  the effective rates at this value can be used to deduce the rates at $R=R_0$.  In other words, based on the value of effective rate at $R/R_0=1/2$, which is denoted as $k_{eff}^{*}$, one can deduce that $1/\tau_D=k_{eff}^*/2^6$.  The estimates for the lifetimes of $D$ based on this assumption are as follows: $0.261\ {\rm ns}$  (Case I, $E_D-E_A=800 \ {\rm cm^{-1}}$);   $0.315\ {\rm ns}$  (Case I, $E_D-E_A=400 \ {\rm cm^{-1}}$); $0.1\ {\rm ns}$  (Case II, $E_D-E_A=800 \ {\rm cm^{-1}}$); $0.099\ {\rm ns}$  (Case II, $E_D-E_A=400 \ {\rm cm^{-1}}$).   Figure 5 shows efficiencies calculated by eq. \ref{eq:eff-1} employing these estimates.   Only the region of  $0.2 < R/R_0 < 0.4$, for which the efficiency is very close to unity, is shown in order to make the discrepancies among different values appreciable.  The  ideal value of efficiency based on the original FRET, eq.  \ref{eq:fret_eff}, is also shown as a reference.  All of these results confirm the trends expected from Fig. 4.   The quantum coherence contributes to the enhancement of efficiency at smaller distances, whereas the non-Markovian bath effects cause moderate reduction of the efficiency at shorter distances.

\begin{figure}
\includegraphics[width=3.5in]{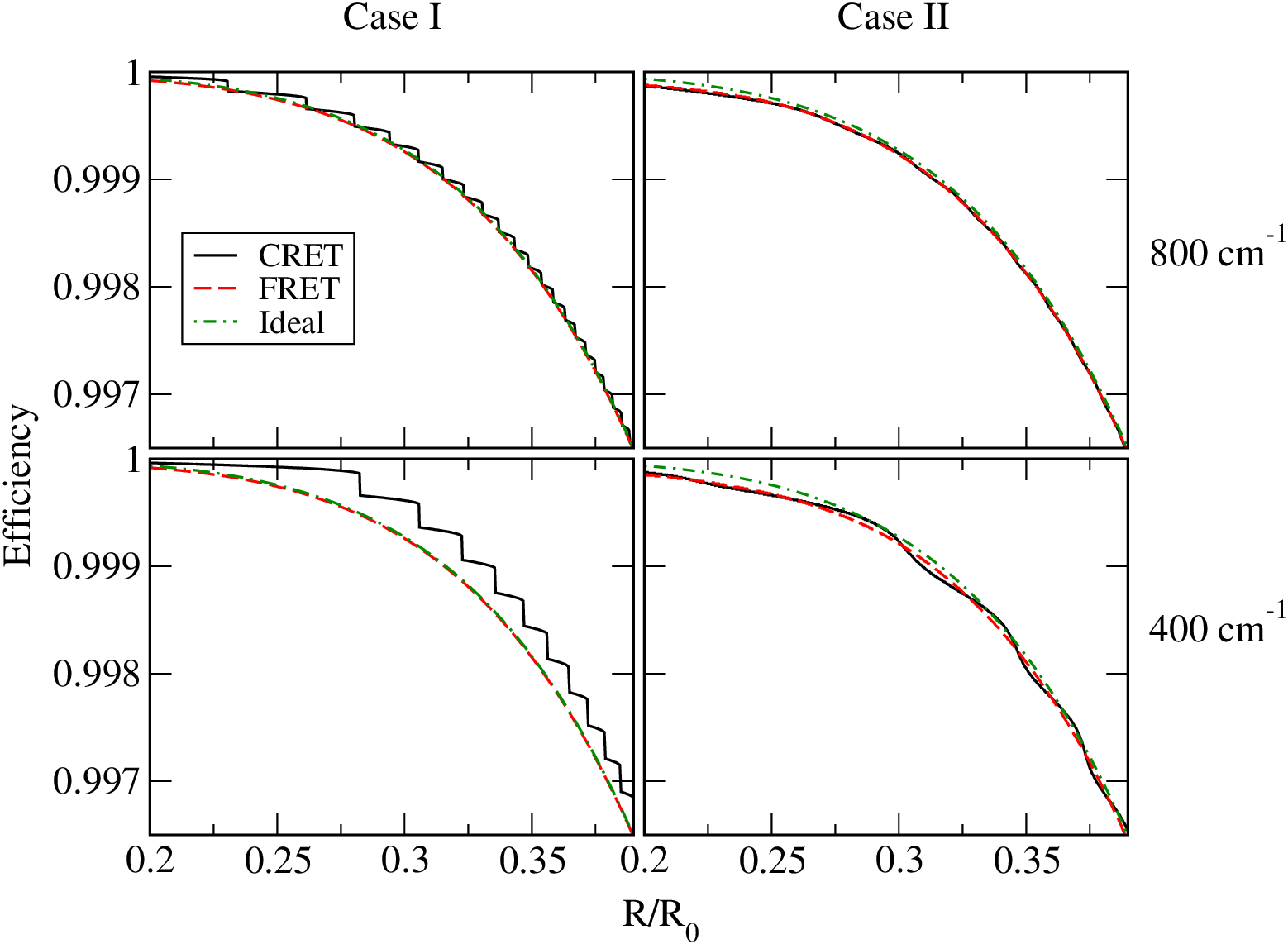}
\caption{Efficiencies calculated by eq. \ref{eq:eff-1} based on both CRET and FRET.  The ideal value of the efficiency based on FRET, eq. \ref{eq:fret_eff}, is shown as reference as well.}
\end{figure}

While the trends in Fig. 5 have significant implications, it is also important to note that the discrepancies shown there are too small to be detected in the presence of typical noises and experimental uncertainties.   Unless extremely high precision measurement is made, larger values of $J_0$ are needed for appreciable changes to occur in the efficiency due to donor-acceptor quantum coherence and non-Markovian bath effects.  This is understandable because the steepest variation of the efficiency occurs due to dynamics with time scales comparable to spontaneous decay lifetimes.  For the choice of  $J_0=5\ {\rm cm^{-1}}$ and the model parameters tested, the estimates for lifetimes were $\sim 0.1-0.3\ {\rm ns}$.  While these are shorter than typical lifetimes of dye molecules, they are still much longer than conventional time scales for quantum coherence and non-Markovian bath effects.   For donors with fast \re{spontaneous} decay rates, either due to additional dark processes or giant oscillator strengths so as to make their lifetimes in the range of hundred femtosecond timescale, detection of the types of discrepancies shown in Fig. 5  \re{is feasible.} 

\section{Conclusion}

While there have been many works examining the effects of quantum coherence and non-Markovian bath on the RET processes,\cite{yang-cp275,ishizaki-jcp130-2,olaya-castro-irpc30,yang-jacs132,jang-wires3,jang-rmp90,jang-njp15,kassal-jpcl4,pachon-pccp14} it is difficult to find systematic studies investigating their effects on the distance dependence of RET.   This is in part due to the fact that distance dependence is a complex issue that involves contributions of many different factors.  In addition, conventional FRET efficiency measurements have employed spectroscopies with poor time resolution in general, which are not appropriate for probing dynamics involving quantum coherence and non-Markovian effects.  Considering recent advances in ultrafast laser spectroscopic techniques and advances in computational capability, it seems possible to overcome such limitations now.    The results provided in this work have significant implications in this regard in that intriguing distance dependences of RET processes can be identified even under simple and well-defined conditions.   

In order to understand the relative importance of factors considered here, investigation of other mechanisms that can affect the RET dynamics at short distances is necessary.  These include the effects of higher order transition multipolar or exchange interaction terms, common or correlated bath modes,\cite{nazir-prl103} quantum vibration of distances and torsional angles,\cite{yang-jcp137} and local field effects,\cite{knox-jpcb106,scholes-jpcb111,wubs-njp18} all of which are expected to have nontrivial distance dependences. Quantum dynamics methods that can address these issues are already established to some extent although continued effort needs to be made to improve their accuracy or applicability.   On the other hand, construction of reliable forms of Hamiltonians amenable for such quantum dynamics calculation seems more challenging at this point because it requires large scale electronic structure calculation including environmental effects.  In addition, more realistic initial conditions\cite{matro-jpc99,ahn-cpl446,das-jcp135} corresponding to actual excitation by pump pulses with finite duration and with uncertainty in excitation energies need to be included for quantitative modeling of spectroscopic data.   These will be 
subjects of future theoretical works. 

\acknowledgements
This work was supported primarily by the National Science Foundation  (CHE-1362926) and in part by the Office of Basic Energy Sciences, Department of Energy (DE-SC0001393).

\appendix
\section{FRET efficiency and radius}
The well-known standard expression for the FRET rate is as follows: 
\be
k_F=\frac{1}{\tau_D} \left (\frac{R_0}{R}\right)^6  ,\label{eq:kf}
\ee
where $\tau_{_D}$ is the lifetime for the spontaneous decay of the excited donor and $R_0$ is the F\"{o}rster radius defined as 
\be
R_0=\left (\frac{9000 (\ln 10)\kappa^2}{128\pi^5 N_A n_r^4}\int d\tilde \nu \frac{f_{_D}(\tilde \nu)\epsilon_{_A}(\tilde \nu)}{\tilde \nu}\right)^{1/6} . \label{eq:r0}
\ee
In the above expression, $f_{_D}(\tilde \nu)$ is the normalized emission spectrum of the excited donor, $\epsilon_{_A}(\tilde \nu)$ is the molar extinction coefficient of the acceptor, $n_r$ is the refractive index, $N_A$ is the Avogadro's number, and $\kappa$ is the orientational factor.  For the case where the transition dipole moments of the donor and the acceptor are defined as 
\ben
&&\boldsymbol{\mu}_{D}=\mu_{D}(\sin\theta_{D}\cos\phi_{D},\sin\theta_{D}\sin\phi_{D},\cos\theta_{D}) , \\
&&\boldsymbol{\mu}_{A}=\mu_{A}(\sin\theta_{A}\cos\phi_{A},\sin\theta_{A}\sin\phi_{A},\cos\theta_{A}) ,
\een
the orientational factor has the following expression:
 \be
\kappa=\sin\theta_{D}\sin\theta_{A}\cos(\phi_{D}-\phi_{A})-2\cos\theta_{D}\cos\theta_{A} . \label{eq:kappa}
 \ee
In real molecular systems in solution phase, $\kappa$ hardly remains constant.  Therefore, it is a common practice to use an average value, $\langle \kappa^2 \rangle$, in eq. \ref{eq:r0}.  However, care should be taken in relying on such expression because the original FRET theory was not derived for the case where $\kappa$ fluctuates or goes through quantum mechanical modulation.  

Equations \ref{eq:kf} and \ref{eq:r0} 
\re{have been popular} because they allow determination of major parameters for the rate based \re{mostly} on experimental data.  However, these expressions are somewhat cumbersome when used along with modern spectroscopic and computational tools and may even cause additional errors if contributions 
\re{from non-radiative decay processes are} included in $\tau_D$.  By simple application of the FGR, it is straightforward to show that 
\be
k_F=\frac{J^2}{2\pi \hbar^2}\int d\omega  L_D(\omega) I_A(\omega) , \label{eq:kf_ang_omega}
\ee
where $L_D(\omega)$ and $I_A(\omega)$ are lineshape functions for the emission of $D$ and the absorption of $A$ defined with respect to the angular frequency $\omega$.  When converted
\re{to} the wavenumber $\tilde \nu=\omega/(2\pi c)$, eq. \ref{eq:kf_ang_omega} can also be expressed as
\be
k_F=4\pi c^2 \tilde J^2 \int d\nu \tilde L_D(\tilde \nu)\tilde I_A(\tilde \nu) ,
\ee 
where $\tilde J=J/(2\pi c\hbar)$, $\tilde L_D(\tilde \nu)=c L_D(2\pi c\tilde \nu)$, and $\tilde I_A(\tilde \nu) =c I_A(2\pi c\tilde \nu)$.
Inserting the expression $J=\mu_D\mu_A\kappa^2/(n_r^2 R^3)$, eq. \ref{eq:kf_ang_omega} can be expressed as
\be
k_F=\frac{\mu_D^2\mu_A^2\kappa^2}{\pi \hbar^2 n_r^4 R^6} \int \tilde d\nu \tilde L_D(\tilde \nu)\tilde I_A(\tilde \nu)  . \label{eq:kf_wn}
\ee
Comparing eq. \ref{eq:kf} with \ref{eq:kf_wn}, one can find the following alternative expression for $R_0$:
\be
R_0=\left (\frac{\tau_D \mu_D^2\mu_A^2\kappa^2}{\pi\hbar^2 n_r^4}\int  \tilde d\nu \tilde L_D(\tilde \nu)\tilde I_A(\tilde \nu) \right)^{1/6}  .\label{eq:r0-1}
\ee
If the lifetime $\tau_D$ is purely due to radiative spontaneous decay, 
\be
\frac{1}{\tau_{_D}}=\frac{2n_r' \mu_D^2 (2\pi c)^4}{3\hbar c^4}\int d\tilde \nu \tilde \nu^3 \tilde L_D(\tilde \nu) \label{eq:tau_d}
\ee
where $n_r'$ is the effective refractive index for the emission including local field effect, which can be different from $n_r$. 
Inserting eq. \ref{eq:tau_d} into eq. \ref{eq:r0-1}, we obtain the following expression:
\be
R_0=\left (\frac{3\mu_A^2\kappa^2}{128 \pi^5 \hbar n_r^4 n_r'} \frac{\int \tilde d\nu \tilde L_D(\tilde \nu)\tilde I_A(\tilde \nu)}{\int d\tilde \nu \tilde \nu^3 \tilde L_D(\tilde \nu)}\right)^{1/6} .
\ee

The value of electronic coupling at $R_0$, as defined by eq. \ref{eq:j_def}, can also be expressed as follows:
\ben
J_0&=&\frac{\mu_D\mu_A\kappa}{n_r^2R_0^3}\nonumber \\
&=&\mu_D\mu_A\left (\frac{128 \pi^5N_A}{9000 (\ln 10) \int d\tilde \nu f_D(\tilde \nu)\epsilon_A(\tilde \nu)/\tilde \nu}\right)^{1/2} \nonumber \\
&=&\mu_D\left (\frac{128 \pi^5 \hbar n_r^4 n_r'}{3}\frac{\int d\tilde \nu \tilde \nu^3 \tilde L_D(\tilde \nu)}{\int \tilde d\nu \tilde L_D(\tilde \nu)\tilde I_A(\tilde \nu)}\right)^{1/2} .
\een

One of the simplest way to derived the FRET efficiency given by eq. \ref{eq:fret_eff} is as follows.  
Consider an ensemble of donors subject to steady irradiation that selectively excites $D$.   In the absence of FRET, the concentration of the excited state donor is denoted as $[D^*]_0(t)$.  Then, its time derivative is determined by  
\be
\frac{d}{dt}[D^*]_0(t)=I_r[D]-\frac{1}{\tau_{_D}}[D^*]_0(t) ,  \label{eq:rate_nofret}
\ee
where \re{$I_r$ is the rate of  incident photons causing excitation, and $[D]$ is the concentration of $D$, the ground state donor.  $I_r$ is} assumed to be small enough to approximate $[D]$ as being constant.
The concentration of the excited stated donor in the presence of FRET is denoted as $[D^*]$.
Then, under the assumption that the reverse process from the excited acceptor can be neglected, 
\be
\frac{d}{dt}[D^*](t)=I_r[D]-\left (\frac{1}{\tau_{_D}}+k_F\right)[D^*](t) . \label{eq:rate_fret}
\ee

Assuming steady state limits where the rate of concentration change of excited state donors becomes virtually zero, the corresponding steady state concentrations can be obtained from eqs. \ref{eq:rate_nofret} and \ref{eq:rate_fret} as follows: 
\ben
&&[D^*]_{0,s}=\frac{I_r[D]}{1/\tau_{_D}} , \\
&&[D^*]_{s}=\frac{I_r[D]}{k_F+1/\tau_{_D}} .
\een
The FRET efficiency can be defined as 
\be
E=1-\frac{[D^*]_{s}}{[D^*]_{0,s}}=\frac{k_F}{k_F+1/\tau_{_D}} . \label{eq:eff-1}
\ee
Inserting the expression for eq. \ref{eq:kf} into the above expression, one can easily show that
\be
E=\frac{\left (R_0/R\right)^6}{1+\left (R_0/R\right)^6}=\frac{1}{1+\left (R/R_0\right)^6} .
\ee

The second expression for $E$ in eq. \ref{eq:eff-1} makes it possible to determine the efficiency in terms of lifetime measurements in the absence and presence of FRET.  Let us define the lifetime of the excited donor in the presence of FRET as $\tau_{_{D,F}}$, which is given by
\be
\frac{1}{\tau_{_{D,F}}}=\frac{1}{\tau_{_D}}+k_F .
\ee
Then, then efficiency can also be expressed as 
\be
 E=1-\frac{\tau_{_{D,F}}}{\tau_{_D}} .
 \ee
\re{The expressions for the efficiency derived in this section are based on the assumption that the spontaneous lifetime of the excited state donor remains the same in the presence of  FRET.  Thus, care should be taken in using them if the new environment for FRET introduces a new non-radiative process or local field corrections.   }  
 
 \section{FRET efficiency in the presence of reverse reaction}
 
The derivation of the FRET efficiency needs to be modified if the reverse transfer from excited acceptor is non-negligible, for which 
 \ben
&&\frac{d}{dt}[D^*](t)=I_r[D]-\left (\frac{1}{\tau_{_D}}+k_F\right)[D^*](t) +k_F^r [A^*](t), \nonumber \\ \label{eq:rate_fret-1} \\
&&\frac{d}{dt}[A^*](t)=k_F[D^*](t)-\left (\frac{1}{\tau_{_A}}+k_F^r\right)[A^*](t) , 
\een
where $k_F^r$ is the FRET rate for the reverse process from $A$ to $D$.
Once again, assuming steady state and weak irradiation, 
\ben
&&[A^*]_s=\frac{k_F[D^*]_s}{k_F^r+1/\tau_{_A}} ,\\
&&[D^*]_s=\frac{I_r[D]}{1/\tau_{_D}+k_F/(k_F^r\tau_{_A} +1)} .
\een
For this case, the efficiency is given by the following expression:
\be
E=\frac{k_F\tau_D}{k_F\tau_D+k_F^r\tau_A+1} .
\ee
Assuming that the reverse process also follows the same FRET process but with different FRET radius $R_0^r$, the efficiency in this case is expressed as 
\be 
E=\frac{1}{1+(R_0^r/R_0)^6+(R/R_0)^6} .
\ee
This is similar to eq. \ref{eq:fret_eff}, but amounts to having a different effective F\"{o}rster radius, $R_0(1+(R_0^r/R_0)^6)^{1/6}$.   The maximum of efficiency is also smaller than unity. 
On the other hand, consider a different excitation process where a short excitation pulse creates $[D^*](0)$ (with $[A^*](0)=0$) at time $t=0$.  Once the excitation pulse becomes inactive, the time dependent concentrations of  $D^*$ and $A^*$ are governed by the following equations:
\be
 \frac{d}{dt}\left ( \begin{array}{l} 
                           $[D*](t)$ \\ $[A*](t)$ 
                           \end{array} \right)  
                           =\left (\begin{array}{cc} -\frac{1}{\tau_{_D}}-k_F & k_F^r \\ k_F & -\frac{1}{\tau_{_A}}-k_F^r \end{array} \right ) \left (\begin{array}{c} $[D*](t)$ \\ $[A*](t)$  \end{array}\right)  .
\ee
The two eigenvalues of the above matrix equation are as follows:
\ben
\lambda_{\pm}&&=k_F\left \{-\frac{1}{2}\left (1+\frac{k_F^r}{k_F}+\frac{\tau_{_D}+\tau_{_A}}{k_F \tau_{_D}\tau_{_A}}\right) \right . \nonumber \\
&&\left . \pm \left [\frac{1}{4}\left ( 1-\frac{k_F^r}{k_F}+\frac{\tau_{_A}-\tau_{_D}}{k_F\tau_D\tau_A}\right )^2+\frac{k_F^r}{k_F}\right]^{1/2}\right\}  .
\een
The corresponding time dependent concentrations are as follows:
\re{
\ben
&&[D^*](t)=\frac{[D^*](0)}{2B}\left [ \left (-A+B \right) e^{\lambda_+ t}+\left (A+B\right ) e^{\lambda_- t} \right ] , \nonumber \\ \\
&&[A^*](t)=\frac{[D^*](0)}{2B}\left [ e^{\lambda_+ t}- e^{\lambda_- t} \right ]  ,
\een
}where
\ben
&&A=\frac{1}{2}\left (1-\frac{k_F^r}{k_F}+\frac{\tau_{_A}-\tau_{_D}}{k_F\tau_{_D}\tau_{_A}} \right) \label{eq:A-exp}  ,\\
&&B=\sqrt{A^2+\frac{k_F^r}{k_F}} .
\een

For the case where $\tau_{_D}=\tau_{_A}=\tau$, the two eigenvalues and $A$ and $B$ become 
\ben
&&\lambda_+=-\frac{1}{\tau} , \\
&&\lambda_-=-k_F-k_F^r-\frac{1}{\tau}  , \\
&&A=\frac{1}{2}\left (1-\frac{k_F^r}{k_F}\right) \label{eq:a-exp} ,\\
&&B=\frac{1}{2}\left (1+\frac{k_F^r}{k_F}\right) \label{eq:b-exp} .
\een
For the case $k_F,k_F^r >> 1/\tau_D,1/\tau_A$,  the following approximations can be used. 
\ben
&&\lambda_+\approx -\frac{k_F\tau_D+k_F^r\tau_A}{(k_F+k_F^r)\tau_D\tau_A} , \\
&&\lambda_-\approx -k_F-k_F^r-\frac{k_F\tau_A+k_F^r\tau_D}{(k_F+k_F^r)\tau_D\tau_A}  , \\
&&B\approx \frac{1}{2}\left (1+\frac{k_F^r}{k_F}\right)+\frac{(k_F-k_F^r)}{2(k_F+k_F^r)}\frac{(\tau_A-\tau_D)}{k_F\tau_A\tau_D} .
\een
On the other hand, the expression for $A$, eq. \ref{eq:A-exp}, does not need further approximation.  
The above expressions provide detailed information on bi-exponential decay rates and their limiting behavior in the presence of RET.  For the case where there is significant delocalization of excitations, similar kinetic equations can be used employing exciton states delocalized between $D$ and $A$.

\re{The two limiting expressions shown above can be used for extracting the FRET rates directly from the bi-exponential fitting of excited donor and acceptor populations.   In the limit of  $\tau_D, \tau_A \rightarrow \infty$, both of these approach the following expressions that can be solved directly from the corresponding kinetic equations:
 \ben 
 &&\lambda_+=0 , \\
 &&\lambda_=-k_F-k_F^r  .
 \een
  with $A$ and $B$ given by eqs. \ref{eq:a-exp} and \ref{eq:b-exp}.  The time dependent concentrations of the excited donor and the acceptor for this case are as follows: 
\ben
&&[D^*](t)=\frac{[D^*](0)}{(1+k_F^r/k_F)}\left [ \frac{k_F^r}{k_F}+ e^{-(k_F+k_F^r)t} \right ]  , \\
&&[A^*](t)=\frac{[D^*](0)}{(1+k_F^r/k_F)}\left [ 1- e^{-(k_F+k_F^r) t} \right ]  .
\een 
These satisfy the detailed balance condition that $[D^*(\infty)]/[A^* (\infty)]=k_F^r/k_F$.  For this case, $\tau_{_{RET}}$ defined for eq. \ref{eq:k_eff} becomes $\tau_{_{RET}}=1/(k_F+k_F^r)$  and the following relation also holds. 
\ben
&&P_D(\tau_{_{RET}})-P_A(\tau_{_{RET}})\frac{P_D(\infty)}{P_A(\infty)}\nonumber \\
&&\hspace{.2in}= \frac{[D^*(\tau_{_{RET}})]}{[D^*(0)]} -\frac{[A^*(\tau_{_{RET}})]}{[D^*(0)]}\frac{k_F^r}{k_F} = \frac{1}{e} .
\een
In combination with the fact that $P_A(\infty)=[A^*](\infty)/[A^*](0)=k_{F}/(k_F+k_F^r)$, the above identity serves as the derivation for eq. \ref{eq:k_eff}.}

\providecommand{\latin}[1]{#1}
\providecommand*\mcitethebibliography{\thebibliography}
\csname @ifundefined\endcsname{endmcitethebibliography}
  {\let\endmcitethebibliography\endthebibliography}{}

\end{document}